\documentclass[superscriptaddress,groupedaddress,nofootnoteinbib,12pt]{article}  

\usepackage{graphicx}  
\usepackage{dcolumn}   
\usepackage{bm}        
\usepackage{jcappub}

\def\be{\begin{equation}}
\def\ee{\end{equation}}
\def\ba{\begin{eqnarray}}
\def\ea{\end{eqnarray}}
\def\beq{\begin{eqnarray}}
\def\eeq{\end{eqnarray}}

\def\mpl{M_{\rm Pl}}

\def\p{{\cal P}}

\def\K{{\cal K}}

\def\L*{{\cal L}_*}
\def\L{\mathcal{L}}
\def\({\left(}
\def\){\right)}

\def\p{\partial}
\def\mn{_{\mu \nu}}
\def\stu{St\"uckelberg }
\def\p{\partial}

\def\<{\langle}
\def\>{\rangle}

 \def\neq {\not\equiv}

\def\cs2{c_{s}^{2}}

 \def\p{\partial}

 \def\be   {\begin{equation}}   \def\ee   {\end{equation}}
 \def\ba   {\begin{array}}      \def\ea   {\end{array}}
 \def\bea  {\begin{eqnarray}}   \def\eea  {\end{eqnarray}}
 \def\bean {\begin{eqnarray*}}  \def\eean {\end{eqnarray*}}

\begin{document}

\title{Cosmological perturbations in Massive Gravity and the Higuchi bound}

\author{Matteo Fasiello}
\author{and Andrew J. Tolley}
\affiliation{Department of Physics, Case Western Reserve University, 10900 Euclid Ave, Cleveland, OH 44106, USA}


\abstract{In de Sitter spacetime there exists an absolute minimum for the mass of a spin-2 field set by the Higuchi bound $m^2 \ge 2H^2$. We generalize this bound to arbitrary spatially flat FRW geometries
in the context of the recently proposed ghost-free models of Massive Gravity with an FRW reference metric, by performing a Hamiltonian analysis for cosmological perturbations. We find that the bound generically indicates that spatially flat FRW solutions in FRW massive gravity, which exhibit a Vainshtein mechanism in the background as required by consistency with observations, imply that the helicity zero mode is a ghost.  In contradistinction to previous works, the tension between the Higuchi bound and the Vainshtein mechanism is equally strong regardless of the equation of state for matter.}

\maketitle


\section{Introduction}
Theories that attempt to modify General Relativity (GR) have a long and rich history. Their study in most recent years has been further motivated by the observational finding on the Supernova data \cite{Perlmutter:1998np,Riess:1998cb} which points to acceleration in the current expansion of the Universe.  If GR is correct, then there must be a \textit{dark energy} density of $\rho\sim 10^{-29}\rm{g/cm^3}$. If this value is due to the cosmological constant $\Lambda$, it will enforce an extremely small number for $\Lambda$,  very far from the value that arises from quantum field theory considerations \cite{Weinberg:1988cp}. On the other hand, if one is willing to modify GR, it has been shown in different \textit{modified gravity} theories \cite{Deffayet:2000uy}-\cite{DeFelice:2010aj} that there exist self-accelerating solutions \cite{Carroll:2003wy,Carroll:2004de} which might explain the current acceleration of the Universe without resorting to dark energy. 
A ubiquitous  issue in theories of gravity, such as the smallness of the cosmological constant, can have different manifestations in different theories. A much desired feature of such theories would be a mechanism by which the smallness of the parameter $\Lambda$ (or whatever parameter takes its place) is technically natural. Quite generally, if setting to zero a small parameter in a theory results in an additional symmetry, it is reasonable to expect that quantum corrections to that parameter will be of the same order of the parameter itself as they are protected by the initial symmetry; if this is the case, the parameter is said technically natural.

Take for example massive gravity (MG): at large distances MG weakens with respect to GR, the potential reads $\sim e^{-m r}/r$. This results in the possibility of an accelerated expansion without dark energy. Observations force one to assume a very small $m$ and again, the fine tuning issue reappears in the ratio $m/M_{\rm Pl}$. Interestingly though, one notices here how the $m=0$ theory, GR, has an additional gauge symmetry that is broken by the massive term and so there is indeed room for a technically natural small value of the parameter $m$. 

Departing from GR comes with a heavy baggage: since the pioneering proposal of Fierz and Pauli \cite{FP}, theories of massive gravity have been plagued with continuity issues \cite{vDVZ1,vDVZ2,Vain}, instabilities (ghosts) \cite{BD}, and with a hierarchy of scales which has made it hard to make sense of the theory \cite{ArkaniHamed:2002sp}. A lot has been done in recent years to make the prospect of a massive theory of gravity more intriguing and, possibly, more predictive. This effort \cite{Kurtrev} culminated \cite{deRham:2010ik,deRham:2010kj} in the formulation of a theory of massive gravity (which we shall refer to as dRGT from now on) which is free of ghosts  at the fully non linear level \cite{Hassan:2011hr,Hassan:2011ea,deRham:2011rn,deRham:2011qq,Mirbabayi:2011aa, Hassan:2012qv,Golovnev:2011aa,Kluson:2011rt}. This theory is endowed with a benevolent hierarchy of scales which neatly splits the linear regime, the non-linear one and the regime where quantum effects must be taken into account.

Having such a young theory at one's disposal, there is no shortage of aspects in need of careful investigation \cite{,Gumrukcuoglu:2011zh,Burrage:2011cr,Alberte:2011ah,Berezhiani:2011mt,Brihaye:2011aa,Vakili:2012tm,Tasinato:2012mf,Baccetti:2012bk}. The more cosmologically inclined might for example opt for a study of realistic cosmological solutions for dRGT, this has recently been done in \cite{massivec}. The analysis we present here shares some features with the work \cite{massivec} in that it points towards the same direction. In this manuscript we study the classical stability of the scalar sector of dRGT theory up to quadratic order in perturbations when the reference metric is taken to be FRW.  We probe several cases according to the background value of the dynamical metric $g_{\mu\nu}$ and the reference metric $f_{\mu\nu}$. Matter content besides the cosmological constant is also considered.  Such a study has been performed in the past for theories which were known to have ghosts \cite{Deser,Grisa}. In this context the so called \textit{Higuchi bound} \cite{Higuchi} on the mass of the graviton was introduced. It is a strong lower bound on $m$, $m^2\ge 2H^2$, and arises from the requirement that the kinetic term for the helicity zero mode, i.e. the scalar cosmological perturbations, is positive definite. Things do not necessarily improve when one includes matter because a similar bound must hold over the different cosmological epochs \cite{Grisa}, as we will see. We show that, even if in dRGT there is more room to accommodate for such a bound (e.g. dRGT has 2 free parameters) and even if we employ at full the freedom on the reference metric, once observations are taken into account the bound remains quite stringent. We hint to a possible resolution in the \textit{Conclusions}. Our final bound is give by
\be
\tilde{m}^2(H) =m^2  \frac{H}{H_0} \left( (3+3\alpha_3+\alpha_4) -2(1+2 \alpha_3+\alpha_4)\frac{H}{H_0}+(\alpha_3+\alpha_4) \frac{H^2}{H_0^2}\right)\ge 2 H^2. \nonumber
\ee 
where $H$ is the Hubble rate of the dynamical metric and $H_0$ that of the reference (non-dynamical) metric, and $\alpha_3$ and $\alpha_4$ are the two free parameters in the dRGT model \cite{deRham:2010kj}. This should be compared with the associated Friedmann equation
\bea
H^2&=&\frac{1}{3 \mpl^2} \rho  -(6+4\alpha_3+\alpha_4)\frac{m^2}{3}+(3+3\alpha_3+\alpha_4)m^2 \frac{H}{H_0}-\nonumber \\ &&(1+2\alpha_3+\alpha_4)m^2 \frac{H^2}{H_0^2} + (\alpha_3+\alpha_4)\frac{m^2}{3}\frac{H^3}{H_0^3}. \nonumber
\eea

In this paper the background metrics studied are both homogeneous and isotropic, a natural next step would then be to consider introducing inhomogeneities at the background level. There is more to support further work in this direction: in \cite{massivec}, it was shown that there exist no truly homogeneous and isotropic cosmological solutions in dRGT theory with a Minkowski reference metric, for the very same reasons that guarantee no Boulware-Deser ghost is present in such a theory.  On the other hand, there are approximate solutions that well describe observations and yet evade this no-go theorem.  We will expand upon this in a forthcoming work \cite{paper3}.

This paper is organized as follows: in the first \textit{Section} we briefly introduce dRGT theory. In \textit{Section 2} we report the details of the theory at second order in perturbation for the scalar sector. In \textit{Section 3} we briefly introduce the Higuchi bound and summarize previous work on the subject. In \textit{Section 4} we present the analysis for the theory when both metrics are de Sitter. In \textit{Section 5} we add matter content  and consider FRW solutions. In the \textit {Conclusions} section we elaborate on our findings and future work.

Although in this work we concentrate on quite specific aspects of the theory, it is important to appreciate that this model of massive gravity seems to enjoy properties and structures that are ubiquitous in current analysis of, for example, alternative models to inflation \cite{galileon1,galileon2,burrage,Bartolo,galileon3}. This theory as a whole also appears to be part of a larger family of massive theories of gravity \cite{Paulos} some of which first emerged in the study of $\rm{AdS}_{3}/\rm{CFT}_2$ correspondence.

\section{ dRGT Massive Gravity}
The theory of massive gravity defined on an arbitrary reference metric $f_{\mu\nu}$ \cite{Hassan:2011tf}  is just a straightforward generalization of the theory proposed in \cite{deRham:2010kj}. The Lagrangian takes the form of Einstein gravity with matter plus a potential that is a scalar function of the two metrics
\be
\mathcal{L}=\frac{\mpl^2}{2}\sqrt{-\, {}^{(4)}\!g}\left(\, {}^{(4)}\!R+{2m^2}\mathcal{U}(g,f)\right)+ \mathcal{L}_M\,. \label{nomatter}
\end{equation}
The most general potential $\mathcal{U}$  that has no ghosts is build out of characteristic polynomials of the eigenvalues of the tensor
\be
\mathcal{K}^{\mu}_{\nu}(g,f)=\delta^{\mu}_{\nu}-\sqrt{g^{\mu \alpha}f_{\alpha \nu}} \, ,
\ee
so that
\begin{equation}
\label{eq:fullU}
\mathcal{U}(g,H)=\mathcal{U}_2+\alpha_3 \, \mathcal{U}_3+\alpha_4 \, \mathcal{U}_4,
\end{equation}
where the $\alpha_n$ are free parameters, and
\begin{eqnarray}
\mathcal{U}_2&=&\frac{1}{2!} \left( [\mathcal{K}]^2-[\mathcal{K}^2]\right),\\
\mathcal{U}_3&=&\frac{1}{3!}\left( [\mathcal{K}]^3-3[\mathcal{K}][\mathcal{K}^2]+2[\mathcal{K}^3]\right),\\
\mathcal{U}_4&=&\frac{1}{4!}\left(  [\mathcal{K}]^4-6[\mathcal{K}^2][\mathcal{K}]^2+8[\mathcal{K}^3][\mathcal{K}]+3[\mathcal{K}^2]^2-6[\mathcal{K}^4]\right) \,,
\end{eqnarray}
where $[\ldots]$ represents the trace of a tensor with respect to the metric $g\mn$.
The absence of ghost for this theory for a Minkowski background metric was shown in the decoupling limit in \cite{deRham:2010gu,deRham:2010ik,deRham:2010kj}, fully non-linearly beyond the decoupling limit in \cite{Hassan:2011hr,Hassan:2011ea}, as well as in the \stu and helicity languages in \cite{deRham:2011rn,deRham:2011qq,Mirbabayi:2011aa, Hassan:2012qv,Golovnev:2011aa,Kluson:2011rt}.
Varying with respect to the metric $g_{\mu\nu}$ we find the equations of motion
\be
G_{\mu\nu}+m^2 X_{\mu\nu} = M_{\rm Pl}^{-2} \, T_{\mu\nu} \, ,
\ee
where 
\beq
\label{eom}
&&X\mn=\K\mn -\mathcal{K}g\mn -(1+\alpha_3) \(\K^2\mn-\K \K\mn+\frac 12 g\mn \([\K]^2-[\K^2]\)\)\\
&&+(\alpha_3+\alpha_4) \(\K\mn^3-\K \K\mn^2+\frac 12 \K\mn \([\K]^2-[\K^2]\)-\frac 16 g\mn \([\K]^3-3 [\K][\K^2]+2[\K^3]\)\)
\notag\,.
\eeq
Using the Bianchi identities, we obtain the following constraint on the metric
\beq
m^2 \nabla^{\mu}X_{\mu\nu}=0.
\label{const}
\eeq
It is clear from the above formula that the subspaces of parameter choices $\alpha_3+\alpha_4=0$ and $\alpha_4=-\alpha_3=1$ are special. We shall see below that the same structure arises in the Higuchi bound. As is well understood, we can introduce \stu fields to recover diffeomorphism invariance, however in what follows we shall work entirely in unitary gauge to make clearer the comparison with the work of \cite{Deser} and \cite{Grisa}.

\section{Action at 2nd order in perturbations}
Our starting point for the two metrics reads:
\bea
 \mathbf{g}_{\mu\nu} = \left(
      \begin{array}{ccc}
        -N^2+N_i^2/a^2 & N_j  \\
       N_j  & a^2\,\delta_{ij}+h_{ij}  \\
      \end{array} \right)\,\,\, ; \qquad \mathbf{f}_{\mu\nu} = \left(
      \begin{array}{ccc}
        -M^2(t) & 0_i  \\
        0_j & b^2(t) \, \delta_{ij}  \\
      \end{array} \right)\,\,\, , \label{metrics}
\eea
where, $f_{\mu\nu}$ being the fixed, reference metric,  is not perturbed\footnote{Giving $f$ full dynamics would take us into the realms of multi-metric ghost-free theories \cite{Pilo,vonStrauss}.}. It is convenient to employ throughout this work the ADM decomposition for the metric and its fluctuations \cite{ADM}. For future convenience we define the ratio of scale factor 
\be
\frac{b}{a}=1+z.
\ee 
If both the background and reference metrics are de Sitter then we would have $\bar g_{\mu\nu}= (1+z)f_{\mu\nu}$ with $z$ time independent. However our formula will be valid in the general time-dependent case. 

The same constraint, coming from the consistency of the Friedmann $\dot{a}$ and acceleration $\ddot{a}$ equations that forbids the existence of FRW solutions for Minkowski reference metric, in the generic FRW case imposes the condition
\be
\left(1+2(1+\alpha_3)z +(\alpha_3+\alpha_4)z^2\right) \left( \frac{b}{a} - \frac{H}{H_0} \right)=0.
\ee
where $H_0$ is the Hubble rate for the reference metric $H_0= \dot{b}/(Mb)$. Although it appears that we could solve this relation with $\left(1+2(1+\alpha_3)z +(\alpha_3+\alpha_4)z^2\right) =0$, it turns out that in this branch of solutions there are fewer  than the full 5 propagating solutions in the gravity sector, and these correspond to singular points in phase space which would be infinitely strongly coupled due to the vanishing of the kinetic term for the helicity zero mode at quadratic order (we shall discuss this further in \cite{paper2}). 

The {\it correct} solution of this relation that gives rise to the full 5 propagating degrees of freedom in the gravity sector is 
\be
\frac{b}{a}= \frac{H}{H_0},
\ee
and it is this solution only that we shall consider in the following. This relationship is crucially important in what follows. It is this relation that ties together the dynamics of the two metrics. This same relationship arises in the bigravity case when both metrics are taken to be dynamical \cite{vonStrauss} (so does the same issue with the pathological branch).

In what follows we shall also set $\mpl=1$ to simplify the calculation, reintroducing it only for clarification when necessary. In compact ADM notation, our general action will look like the following:
\bea
S=\int d^4 x \Big[\pi^{ij} \dot{g}_{ij} + \Pi \dot{\Phi}+N \mathcal{E}^{0}+N_i \mathcal{E}^{i}   \Big]  + S_{m} (N,N_i, N_i^2,...)   \label{prima}
\eea
 where $\mathcal{E}^0= \mathcal{R}^0-\mathcal{T}^0$ and $\mathcal{E}^i= \mathcal{R}^i-\mathcal{T}^i$ with\footnote{Note that in ADM notation $\sqrt{g}=\sqrt{{}^{(3)}\!g}$ and $R={}^{(3)}\!R$ as opposed to$\sqrt{-{}^{(4)}\!g}$, ${}^{(4)}\! R$. We follow \cite{Grisa} .}:
\bea
\mathcal{R}^{0}=\sqrt{g}R +\frac{\pi^{ij} \pi^{lm}}{\sqrt{g}}\left (\frac{1}{2} g_{ij}g_{lm}-g_{il}g_{jm} \right); \qquad \mathcal{R}^i = 2 \sqrt{g} D_j \left( \frac{\pi^{ij}}{\sqrt{g}}\right); \nonumber \\
\mathcal{T}^0 =\sqrt{g}\left( \frac{1}{2}g^{ij} \p_i \Phi \p_j \Phi+ V(\Phi) \right) +\frac{\Pi^2}{2\sqrt{g}}; \qquad \,\,\, \mathcal{T}^i =\Pi \p^i \Phi .\qquad \qquad \,\,
\eea
 The lapse and shift functions are auxiliary variables. $N_i$ can be solved for and will drop out of the action while, $N$ will enforce a constraint. With fluctuations in mind, we specify that:
\bea
\pi^{ij}=\bar\pi^{ij} +p^{ij}; \quad N=1+n;\quad \Phi=\phi_0+ \varphi;\quad \Pi=\bar\Pi + \pi.
\eea
 Note that we have set $N=1$ in the background which we ensure by appropriate choice of coordinates for the reference metric, this in general implies $M \neq 1$.
No confusion should arise from the common labeling $\pi^{i}_{i}=\pi$ as opposed to the fluctuation of the conjugate momentum $\pi$ of the scalar $\Phi$ as what we refer to will be quite clear from the context.   

As mentioned, we will concentrate here on the instability of the scalar sector of the theory, since this is where the Higuchi bound is known to arise; to this aim, it is very convenient to employ the following tensor decomposition for  $p^{ij}$ and for the metric fluctuation $h_{ij}$:
\bea
h_{ij}= h^{T\,\,t}_{ij} + \p_{\left( \right. i} h_{\left. j\right)}^t +\frac{1}{2}\Big[ \delta_{ij}-\frac{\p_i \p_j}{\p_k^2}  \Big] h^t + \frac{\p_i\p_j}{\p_k^2}h^l
\eea
Equipped with these conventions and notation, we now write down the most general expression for the massive gravity contribution, it reads: 
\beq
 &&2 m^2 \Big[\left(h_l^2+\frac{h_t^2}{2}\right)d_2  +(h_l+h_t)^2 c_2+e_2 (h_l+h_t) \,n+f_2 N_i^2\nonumber \\&&-\alpha_3 \left((h_l+h_t)^2 c_3+\left(h_l^2+\frac{h_t^2}{2}\right) d_3+(h_l+h_t) n\, e_3 +f_3 N_i^2\right)\nonumber \\&&+\alpha_4 \left((h_l+h_t)^2 c_4+\left(h_l^2+\frac{h_t^2}{2}\right) d_4+(h_l+h_t) n\, e_4+f_4 N_i^2\right)\Big]    \, ,        \label{semiesplicita}
\eeq
 where $c_{n}, d_{n}, e_{n},f_{n} $'s are in general functions of $a,b,M$ as of Eq.~(\ref{metrics}) (see \textit{Appendix A} for an explicit expression for the functions above).
\subsection{The Higuchi bound, an example}

At this stage it seems apt to introduce the stability condition we are after, which will determine the Higuchi bound. In order to do so, we restrict ourselves to a very simple specific case encompassed by the theory above (see \cite{Deser} for a detailed treatment):\\
$\bullet$ no matter content\\
$\bullet$ $\alpha_3=0=\alpha_4$\\
$\bullet$ the reference metric $f_{\mu\nu}$ is set equal to the background value of $g_{\mu\nu}$, i.e. $f_{\mu\nu}=\bar{g}_{\mu\nu}\Leftrightarrow z=0$, so no background contribution from the MG action\footnote{In other words, we are projecting our theory into the region of the parameter space that gives back Fierz-Pauli \cite{FP}.}. 

As we will do for the general case, we write the resulting canonical action in this form:
\bea
L= {\bf p}^{T} \cdot \dot {\bf q}-\frac{1}{2} \,{\bf p}^{T} \cdot K_{(,)} \cdot {\bf p}-\frac{1}{2} \,{\bf q}^{T} \cdot M_{(,)} \cdot {\bf q}- \,{\bf p}^{T} \cdot V_{(,)} \cdot {\bf q}            . \label{forma}
\eea
The stability conditions can be read off the matrix $K$ and $M$, they should be positive definite for the system to be stable. As will be later explained, we will  focus here on what is now the matrix $M$ and disregard the so called \textit{gradient} instability. In the simplest case, when the three conditions above are met, the Lagrangian for the helicity zero mode is \cite{Deser}: 
\beq
S_0=\int d^4 x \Big[\pi^{ij} \dot{g}_{ij} +N \mathcal{E}^{0}+N_i \mathcal{E}^{i}   \Big]  -\frac{m^2}{4 a}\left ( h^{ij}h_{ij}- h^2_{ii} -2a^2 N_i^2 -4a^2 n h_{ii} \right)   \, ,\label{des1}
\eeq
 where the matter $\Phi$ has been put to zero also in $\mathcal{E}^0,\mathcal{E}^1$ and the MG Lagrangian starts at second order in perturbations. The background equations of motion simply read $3H^2=\Lambda, \,\,\, \dot H =0$. 

 We spare the reader the simple albeit tedious algebra but list the steps required to reduce the Lagrangian in (\ref{des1}) to a convenient version of Eq.~(\ref{forma}):\\
\newline 
1) Solve for $N_i$ and plug back in the Lagrangian;\\
2) Employ the field redefinitions: $h_{ij}\rightarrow a^{1/2} h_{ij},\,\, \,\,p^{ij}\rightarrow a^{-1/2} p^{ij},\, \,\,\, n\rightarrow a^{-3/2} n $;\\
3) Solve for the variable $p^t$ (no loss of generality);\\
4) Perform the field redefinitions $p^l \rightarrow p_0 +\frac{\nu^2}{4H} h^t,\, \,\,\, h^l\rightarrow q_0 +\frac{1}{2}h^t,\,\,\,\,\nu^2=m^2-2H^2$ ;\\
5) Solve for $h^t$ by its algebraic equation;\\
6) Finally, use the background e.o.m. and employ the field redefinitions below:
\beq 
p\rightarrow p_0 + H \Big[q_0 +\frac{2H}{\nu^2 m^2} \left( -2\nabla^2 +3\nu^2 -3H^2 \right)p_0  \Big], \qquad q_0\rightarrow q_0 +\frac{2H}{\nu^2 m^2} \left( -2\nabla^2 +3\nu^2 -3H^2 \right)p_0 \,.\nonumber \\
\eeq
 The result is:
\beq 
L=p_0 \dot q_0 -\Big[ \frac{1}{2} \left(\frac{\nu^2 m^2}{12 H^2} \right )q_0^2 +\frac{1}{2}\left( \frac{12 H^2 }{\nu^2 m^2 } \right)p_0 \big[-\nabla^2 +m^2 -9/4\big] p_0  \Big].
\eeq
 Having switched off the scalar $\Phi$ in this example, the canonical variables are just the couple $(p_0,q_0)$; the stability condition can be read directly from the sign of the $q_0^2$ coefficient, that is:
\beq
\nu^2\ge 0 \Leftrightarrow m^2\ge 2H^2
\eeq
 This is the Higuchi bound \cite{Higuchi} on $m$ in this set up. One is of course well aware that, over the different cosmological epochs, it is not allowed to assume $\dot H=0$. Indeed, the bound as it presents itself here, serves as a warning that things need to be checked in more general scenarios, away from de Sitter \cite{Grisa}.
 
We will perform steps similar to the ones mentioned above for the general theory but a quick route to Higuchi's result is worth it for clarifying the physical importance of the bound and how it comes about. Now, observationally, a lower limit on $m$ is close to the last thing one worries about in MG. Quite on the contrary, the importance of identifying an upper bound is intuitively clear, after all, GR is known to work quite well.  

\subsection{Higuchi VS Vainshtein}
In the $m\rightarrow 0$ limit MG should reproduce GR. This does not happen quite so easily because the two theories have a different number of degrees of freedom. Vainshtein was the first to propose \cite{Vain} an expansion parameter for MG solutions which makes it clear how inside the so called \textit{Vainshtein radius} non-linearities play an important role in the theory.  It has been proven that non-linearities help hide the additional degrees of freedom of MG and restore continuity with GR \cite{Deffayet:2001uk,Babichev:2009us,Chkareuli:2011te}.

A heuristic way to obtain the Vainshtein radius is the following. Consider a generic background equation for MG, schematically, it will be of the type:
\[
R_{\mu\nu}+m^2 h_{\mu\nu}+...\sim \frac{1}{\mpl^2}T_{\mu\nu} 
\]
The linear to non-linear regime transition happens when the fluctuation $h_{\mu\nu}\sim 1$ so that $R_{abcd}\sim m^2$. We also know that  $R_{abcd} \sim \nabla^2 \phi$ and $\phi \sim \frac{G M}{r}$. Therefore, $ R_{abcd}\sim\frac{GM}{r^3}\sim m^2 $ and finally:
\beq
r_V=\left(\frac{M}{m^2\mpl^2} \right)^{1/3} \, .
\eeq
 This is the celebrated Vainshtein radius. It is now clear that too large an $m$ would be excluded by tests on the validity of GR in the solar system and beyond. We are inside $r_V$ then, and the mildest condition to impose on  $m$ would be for it to give a negligible contribution to the Friedmann equation of the theory. Since a cosmological constant contribution is allowed by data, the bound is best placed on $\dot{H}$ which for most of cosmic history must be consistent with GR.
In other words consistency with cosmological observations for the period before dark energy domination requires
\be
\label{vain10}
m^2 \frac{d}{dt} \left[ (3+3\alpha_3+\alpha_4) \frac{H}{H_0}-(1+2\alpha_3+\alpha_4) \frac{H^2}{H_0^2} + (\alpha_3+\alpha_4)\frac{1}{3}\frac{H^3}{H_0^3}\right] \ll H \dot{H} .
\ee
On the other hand we expect $\frac{1}{H} \frac{d}{dt} \ln (H/H_0) \sim {\cal O}(1)$. To see this, suppose that the reference metric describes a power-law FRW geometry. This would correspond to  $H_0 = A \, b^n $ for some constants $A$ and $n$. At the same time as we will see below we require $\frac{b}{a} = \frac{H}{H_0}$,
which implies
\be
\frac{b}{a} =\left(\frac{H}{Aa^n} \right)^{1/(1+n)}
\ee
and so if in turn $a(t) \sim t^p$ then
\be
\frac{1}{H} \frac{d}{dt} \ln (H/H_0) = -\left( \frac{pn+1}{p(n+1)}\right) \sim {\cal O}(1)
\ee
For a given epoch of cosmological expansion (\ref{vain10}) typically amounts to a condition of the form 
\be
m^2  \left[ (3+3\alpha_3+\alpha_4) \frac{H}{H_0}-(1+2\alpha_3+\alpha_4) \frac{H^2}{H_0^2} + (\alpha_3+\alpha_4)\frac{1}{3}\frac{H^3}{H_0^3}\right] \ll H^2
\ee
which approximately (assuming one power of $H/H_0$ dominates) is
\be
\tilde{m}^2 \ll {\cal O}(1) H^2
\ee 
where $\tilde{m}^2$ is the effective mass arising in the Higuchi bound (see below). It is this relation which creates tension with the Higuchi bound $\tilde{m}^2 \ge 2H^2$.  We can of course imagine evading the Vainshtein bound for a specific period of cosmic evolution. For instance is $b$ is chosen to behave as in a matter dominated universe, then during the period for which $a$ is also matter dominated, the ratio $H/H_0$ will remain constant, and so the mass term contributes nothing but an overall cosmological constant to the Friedman equation, which may be absorbed into any existing contribution. However, without modifying $b$, the period of radiation domination will necessarily be modified from GR predictions. We will disregard such extreme tunings.

We see now that two different checks on massive gravity, one which originates from studying the classical stability of the theory, the other from the necessity to make sense of observations and recover GR at least at early cosmic times, come together and force contradictory bounds on $m$. The order one numerical coefficients might appear to leave some room for maneuvering, it is therefore a good time for a more detailed analysis.  How this tension might be resolved in ghost-free theories of MG will be the subject of the following sections and of a follow-up work \cite{paper2,paper3}.
 
\section{Higuchi bound: dS on dS}

In the example above, several assumptions have been made, chief among which the use of FP theory, one which has long been known to have instabilities, ghosts.
We now proceed to perform an analogous analysis employing the dRGT theory of massive gravity in full. Let us stress again that here we are dealing with a ghost-free model at fully non-linear level endowed with a clear-cut and convenient hierarchy of scales. For the moment, we do not add matter content but make full use of the parameter space of the theory, including the free parameters $\alpha_3,\alpha_4$. What we will learn for this case will be qualitatively true for the most general one, i.e. the effect of the two parameters  does not change the qualitative picture of the Higuchi-Vainshtein tension. 

We start with the Friedmann equation for the dRGT theory, without matter, according to Eq.~(\ref{nomatter}):
\beq
H^2 =\frac{1}{3} \Lambda+ m^2(z-z^2)-\alpha_3 \,m^2(z^2-z^3/3)+\frac{1}{3}\alpha_4\, m^2 z^3  \label{wo}
\eeq
where $1+z=b/a$. This is just a special case of the general Friedmann equation for an FRW reference metric
\be
H^2 =\frac{1}{3M_{\rm Pl}^2} \rho+ m^2(z-z^2)-\alpha_3 \,m^2(z^2-z^3/3)+\frac{1}{3}\alpha_4\, m^2 z^3 ,
\ee
which can easily be determined from the equation of motion (\ref{eom}).
Consistency of the Friedmann equation and the acceleration equation $\ddot{a}$ (determined from (\ref{eom})) implies the relation\footnote{We disregard as pathological the branch of solutions for which $\left(1+2(1+\alpha_3)z +(\alpha_3+\alpha_4)z^2\right) =0$ \cite{paper2}).}
\be
\left(1+2(1+\alpha_3)z +(\alpha_3+\alpha_4)z^2\right) \left(\frac{b}{a}- \frac{H}{H_0} \right),
\ee
and so
\be
\frac{b}{a}= 1+z=\frac{H}{H_0}.
\ee

Putting this together the general Friedmann equation is (applicable for an arbitrary FRW reference metric)
\bea
H^2&=&\frac{1}{3 \mpl^2} \rho  -(6+4\alpha_3+\alpha_4)\frac{m^2}{3}+(3+3\alpha_3+\alpha_4)m^2 \frac{H}{H_0}-\nonumber \\ &&(1+2\alpha_3+\alpha_4)m^2 \frac{H^2}{H_0^2} + (\alpha_3+\alpha_4)\frac{m^2}{3}\frac{H^3}{H_0^3}. 
\eea

We will use the Friedmann equation above in determining the Higuchi bound but at this stage we can already tell, because of the Vainshtein screening mechanism we introduced above, that the following inequality must hold before the period of dark energy domination:
\beq
m^2\Big[ (z-z^2)-\alpha_3 \,(z^2-z^3/3)+\frac{1}{3}\alpha_4\,  z^3\Big] \lesssim H^2
\eeq
or more precisely
\beq
m^2 \frac{d}{dt}\Big[ (z-z^2)-\alpha_3 \,(z^2-z^3/3)+\frac{1}{3}\alpha_4\,  z^3\Big] \lesssim H \dot{H}
\eeq
We now proceed to give a detailed account on how to obtain the Higuchi bound in this case\footnote{There is a faster way to attain the Higuchi bound in this case, simply introducing a dressed mass $\tilde{m}^2$ that arises from looking at fluctuations about the de Sitter metric utilizing the de Sitter symmetry, and the using the representation theory statement that $\tilde{m}^2>2 H^2$. It is instructive and convenient to spell out at this stage the more general procedure, where we cannot make use of the de Sitter symmetry, doing it directly for the most general theory would sacrifice clarity.}. The explicit version of the calculation for the most general case will only be hinted at in the text of next section, while details will be reported in \textit{Appendix B}.  The explicit expression for the Lagrangian at second order in perturbations for dRGT without matter is: 
\beq
&&\frac{\left(\p_i h_t\right)^2}{8 a^3}-\frac{3 H^2 h_l h_t}{a}-\frac{3 H^2 h_t^2}{4 a}+\frac{9 H^2 (h_l+h_t)^2}{4 a}-\frac{4 H^2 \left(h_l^2+\frac{h_t^2}{2}\right)}{a}+\dot{h}_{ij} p^{ij}-H (h_l+h_t) (p_l+p_t)\nonumber \\
&&
+\frac{1}{2} a (p_l+p_t)^2+2 H \left(h_l p_l+\frac{h_t p_t}{2}\right)-a \left(p_l^2+\frac{p_t^2}{2}\right)+n \left(-\frac{\p_i^2h_t}{a}-2 a^2 H (p_l+p_t)  \right. \nonumber \\ &&\left. +a (h_l+h_t) \left(H^2-\Lambda \right)\right)+\frac{h_l^2 \Lambda }{4 a}-\frac{h_l h_t \Lambda }{2 a}-2 \left(\frac{(h_l+h_t)^2}{8 a}-\frac{h_l^2+\frac{h_t^2}{2}}{4 a}\right) \Lambda -a (h_l+h_t) n \,  \Lambda\nonumber  \\ &&+2 \left(\p_i p_l-\frac{\p_i h_l H}{a}+\frac{\p_i h_t H}{a}\right) N_i+2 m^2 \left(\left(h_l^2+\frac{h_t^2}{2}\right) d_2+(h_l+h_t)^2 c_2+(h_l+h_t) n e_2 \right. \nonumber \\ && \left. 
+f_2 N_i^2-\alpha_3 \left((h_l+h_t)^2 c_3+(h_l+h_t)n e_3+\left(h_l^2+\frac{h_t^2}{2}\right) d_3+f_3 N_i^2\right) \right.\nonumber \\ &&
\left. +\alpha _4 \left((h_l+h_t)^2 c_4+(h_l+h_t)n e_4+\left(h_l^2+\frac{h_t^2}{2}\right) d_4+f_4 N_i^2\right)\right)
\eeq
 We refer the reader to \textit{Appendix A} for the explicit expression of the functions $c_n, d_n, e_n, f_n$ with the reminder that, in this case, $M=1+z$:  we are in dS space. The next step is to solve for $N_i$, to obtain:
\beq
N_i=\frac{-a \,\p_i\, p_l+\p_i \,h_l H-\p_i\, h_t H}{2 a m^2 (f_2-f_3\, \alpha_3+f_4\, \alpha_4)}
\eeq
Plugging this back in the action, we proceed to a convenient redefinitions of variables: 
\beq
h_{ij}\rightarrow a^{1/2} h_{ij},\quad p^{ij}\rightarrow a^{-1/2}p^{ij},\quad n\rightarrow  a^{-3/2}n . \label{redef1}
\eeq
 This will be complemented by the redefinitions on the matter scalar mode in the most general case. We are now ready to eliminate $n$, which acts as a Lagrangian multiplier, and use the constraint to solves for $p_t$:
\beq
&&p_t= -p_l -\frac{\p_i^2 h_t}{2H a^2}+ \frac{\left(h_t+h_l\right)}{2H} \left(2m^2 (e_2-\alpha_3 e_3 +\alpha_4 e_4 ) a^{-1}+ (H^2-2 \Lambda)   \right) = \nonumber\\
&& =  -p_l -\frac{\p_i^2 h_t}{2H a^2}+ \frac{\left(h_t+h_l\right) \nu^2}{2H};\qquad \nu^2 \equiv 2m^2 (e_2-\alpha_3 e_3 +\alpha_4 e_4 ) a^{-1}+ (H^2-2 \Lambda).\nonumber \\
\eeq
 It is now convenient to perform a change of variables \cite{Deser}:
\beq
p_l\rightarrow p_0 +\frac{\nu^2}{4H}h_t,\quad h_l\rightarrow q_0 +\frac{h_t}{2}, 
\eeq
 after which one solves the algebraic equation for $h_t$. The equation is easily solved in coordinate space because, as one can easily check, the coefficient of the $\nabla^2 h_t$ term in the action vanishes at this stage upon using the equation of motion. Finally, in order to put the action in a form where there are no mixed terms of the type $p\cdot  q$ (this is easily done in this case)  we perform the canonical transformation:
\beq
p_0 \rightarrow P+H \left(Q+ A\cdot  P \right),\qquad q_0\rightarrow Q+ A\cdot  P  \label{canonical1}
\eeq
 where $A$ is determined by requiring indeed that the coefficient of $Q\cdot P$ is zero.  We refer to \textit{Appendix A} for the explicit expression. At this point the action will have the simpler form:
\beq
P \dot Q -\Big[ \frac{1}{2}\, \alpha \,P^2 +\frac{1}{2}\,\beta\, Q^2 \Big] .
\eeq
 The last step of the analysis in then to require $\alpha\ge 0$. Explicitly:
\beq
&&m^2  (1+z) (-1+z (2-\alpha_3 (-2+z)-\alpha_4 z)) \times \nonumber \\ &&\left(2 H^2+m^2 (1+z) (-1+z (2-\alpha_3 (-2+z)-\alpha_4 z))\right) \ge0 
\eeq
 There is indeed now an \textit{effective mass}, 
\be
\tilde{m}^2=m^2  (1+z) (1-z (2-\alpha_3 (-2+z)-\alpha_4 z)).
\ee
 One realizes the Higuchi bound is formally identical to the one we obtained in the simplest case:
\beq
\tilde{m}^2\left( \tilde{m}^2-2H^2  \right)\ge 0.
\eeq
Although this bound is easily satisfied if $\tilde{m}^2<0$ this is well-known to give instabilities in the vector sector of the theory since the kinetic term for the vector modes is proportional to $\tilde{m}^2$. Consequently the real Higuchi bound is
\be
\tilde{m}^2 \ge 2H^2.
\ee
In order to get an intuition for the Higuchi mechanism in this case, let us temporarily set $\alpha_3, \alpha_4$ to zero. The form of the bound now is:
\beq
&& m^2(1-z-2z)-2H^2 \ge0  \Leftrightarrow \nonumber \\ &&
\left(3\frac{H} {H_0}-2\frac{H^2}{H_0^2}\right) m^2-2 H^2 \ge 0 \nonumber
\eeq
Now it is clear what one gains when considering a reference metric which is other than the background value of the dynamical metric $g_{\mu\nu}$: the effective mass can be positive while, at the same time, the initial ``bare'' mass $m^2$ is allowed to be negative. Below we present the conditions that need be satisfied for the stability of the Hamiltonian. We split them in two convenient subsets, two branches for the value of $H$, assuming that the effective mass is positive for both . We obtain:
\beq
\left\{
    \begin{array}{rl}
 0<H<\frac{3 H_0}{2}  \,\, \,\,\, \\
\\
  m^2\ge\frac{2 H H_0^2}{3 H_0-2H} \,\,\, \\
\\
\tilde{m}^2>0\,\,\,\,\,\, \qquad \qquad
      \end{array} \right.
\qquad \qquad \qquad \qquad \qquad \quad 
\left\{
    \begin{array}{rl}
     H > 3 \frac{H_0}{2}\qquad   \quad \,   \\
\\
       m^2 < -\frac{2 H H_0^2}{2 H - 3 H_0} \quad \\
\\
\tilde{m}^2>0\,\,\,\,\,\, \qquad \qquad
      \end{array} \right.
\eeq
 where $\tilde{m}^2= \left(3H/H_0-2H^2/H_0^2\right)m^2$. The branch on the left is the one which  represents a continuous deformation of the usual case\footnote{It is important to note here that there is no a priori reason to assume $f_{\mu\nu}=\bar{g}_{\mu\nu}$, other than sheer simplicity. }, for which $z=1 \Leftrightarrow H=H_0$. In such a branch, we recover the by now familiar $m^2>2H^2$ in the $H\rightarrow H_0$ limit. This solution will share the same properties  we elaborated upon in the first example: the bound itself is quite stringent and there is a strong Higuchi-Vainshtein tension (see Eq.~\ref{hig_vai} below).  The other branch is new: it lives in a different region of the parameter space where $H> 3/2 H_0$, allows for a positive effective mass while keeping the bare mass small and negative. Let us rewrite the second condition:
\beq
|m^2|\frac{H}{H_0}\left(2\frac{H}{H_0} -3\right)\ge  2H^2, \,\,\,\,\,\,\,{\rm for}\,\, H\gg H_0\,\Rightarrow\,\, \frac{|m^2|}{H_0^2}>1 \label{Higuchiwo}
\eeq
 So, considering the freedom on the value of $H_0$, the Higuchi bound, by itself, is not that stringent anymore. It can be easily satisfied. The only caveat would be that $H$ must always be larger than $H_0$, which fixes the value of $H_0$ as smaller than the current value of the Hubble constant.  This analysis must be complemented with observations, i.e. MG must give very subleading contributions to the Friedmann equation. Let us rewrite Eq.~(\ref{wo}), again setting $\alpha_3=0=\alpha_4$, in this form: 
\beq
\left(H+\frac{3|m^2|}{2H_0 (1-\frac{|m^2|}{H_0^2})} \right )=\left(\frac{\frac{\Lambda}{3} +2|m^2| +\frac{9m^4}{4H_0^2(1-\frac{|m^2|}{H_0^2})}}{1-\frac{|m^2|}{H_0^2}} \right)^{1/2}.
\eeq
Clearly, we want to recover the $H=\sqrt{\rho/3}$ as $m\rightarrow 0$. By looking at the main denominator on the RHS above, one can see that the condition $|m^2|/H_0^2 <1$ must be imposed and this conflicts with the Higuchi bound in its latest disguise as of Eq.~(\ref{Higuchiwo}). 

By using dRGT theory and its cherished properties, we found a new branch where $H$ can live which is devoid of the typically stringent Higuchi bound. As soon as we focus on the Friedmann equation we realize this branch is not viable anymore. Things do not change qualitatively when switching on $\alpha_3,\alpha_4$, but we report the result below for completeness. Including matter is the next conceptual step, which we introduce in next section.
 The Higuchi bound with all parameters in use reads: 
\beq
m^2(1+z) (1-z (2-\alpha_3 (-2+z)-\alpha_4 z))\ge {2 H^2}{}, \label{higu1}
\eeq
that is, 
\be
\tilde{m}^2(H) =m^2  \frac{H}{H_0} \left( (3+3\alpha_3+\alpha_4) -2(1+2 \alpha_3+\alpha_4)\frac{H}{H_0}+(\alpha_3+\alpha_4) \frac{H^2}{H_0^2}\right) \ge 2H^2 .
\ee
This of course, must be combined with the requirement due to the Vainshtein mechanism, which is obtained straightforwardly from the Friedmann equation :
\beq
m^2\left( 3z - 3z^2 -3\alpha_3 z^2 +\alpha_3 z^3 +\alpha_4 z^3\right) \ll  3H^2 \label{vain1}
\eeq

 Combining the last 2 equations together one obtains: 
\beq
\frac{1-z-2z^2 -2\alpha_3 z -\alpha_3 z^2 +\alpha_4 z^2 +(\alpha_3  +\alpha_4) z^3}{3z - 3z^2 -3\alpha_3 z^2 +(\alpha_3 +\alpha_4) z^3 }\gg1 . \label{hig_vai}
\eeq
There exist specific values of $\alpha_3, \alpha_4, z$ for which this inequality is indeed satisfied. On the other hand, this would  require some extreme tuning on the theory while we have yet to consider the fact that, eventually,  $H$ and so consequently $z$ is supposed to wildly vary during cosmological epochs and this alone might make seemingly special values of $\alpha_3,\alpha_4$ unacceptable.  It is interesting to note that both polynomials in the numerator and denominator are cubic unless $\alpha_3+\alpha_4=0$ in which case they are both quadratic. If in addition $\alpha_3=-1$ then both polynomials are linear. Because of this fact that regardless of the choice of $\alpha_3$ and $\alpha_4$, the polynomials entering the Higuchi bound and the Friedmann equation are the same order, playing with these parameters does not remove the Higuchi-Vainshtein tension.

Let us stress again that, despite using a similar analysis, the result  obtained here differs from the ones in \cite{Deser,Grisa} (as well as from more recent work in \cite{Berkhahn:2010hc,Berkhahn:2011hb}) in more than one, crucial, ways. First of all, the MG theory itself is a different one: the place of Fierz-Pauli MG is taken up by dRGT massive gravity for the reasons on which we expanded upon before. The second important point is that the reference metric $f_{\mu\nu}$ is here allowed to be other than the background value of the dynamical metric $g_{\mu\nu}$, i.e. $f_{\mu\nu}\not= \bar{g}_{\mu\nu}$, the difference is parametrized by the time dependent function $z$.

Before one can claim that the Higuchi-Vainshtein tension is here to stay, matter should be added  to the picture. This is what we do below. From the scalar sector perspective, this corresponds to adding another scalar mode and could, in principle, help relax the Higuchi bound and therefore the Higuchi-Vainshtein tension.

\section{FRW on FRW}
Consider now dRGT theory of massive gravity with FRW reference metric, and add matter content to it. We consider a simple expression for the matter contribution  to the scalar sector, namely a scale field with an arbitrary potential, but we do not anticipate that the conceptual results will part from ours in more general scenarios. The idea is that our model for the scalar can mimic a generic equation of state, and so this should suffice for the instability analysis.

As we have seen in the dS case, the Higuchi-Vainshtein tension forces $m$ into small interval of values.
It is then crucial to investigate what happens away from de Sitter, when matter content is included and so the theory can describe the dynamics over the different cosmological epochs. The full Lagrangian now reads:
\beq
L=L_{dRGT}+\int d^4 x \sqrt{-{}^{(4)}\!g}\left(-\frac{1}{2}g^{\mu\nu}\p_{\mu}\Phi\p_{\nu}\Phi + V(\Phi) \right) 
\eeq 
This Lagrangian has the form of (\ref{prima}). We have shown above that  employing the freedom on the parameters $\alpha_3,\alpha_4$ of the theory does not change the qualitative picture of the stability analysis. This allows us to set $\alpha_4=0=\alpha_3$, which benefits the algebra as well. All of the following calculation steps may be straightforwardly generalized to the case where $\alpha_3 $ and $\alpha_4$ are nonzero. In a future work we shall give a simpler derivation which is applicable in the general case and confirms the following results \cite{paper2}.

We briefly sketch the route to the Higuchi bound, which closely mimics the simpler one above. The main difference is that we have an additional scalar mode and therefore Eq.~(\ref{forma}) is to be intended with matrices as oppose to bare numbers. As a consequence, the calculation is a bit more involved and for simplicity we do not show all the explicit steps in the text. We refer the reader to \textit{Appendix B} for details on all the initial matrices entries and more. 

The equations of motion for the full theory read:
\beq
&&H^2 =m^2 (z-z^2)+\frac{\bar {\pi}^2}{12} +\frac{V}{6} ; \nonumber  \\
&&\dot H =-\frac{\bar {\pi}^2}{4}-\frac{m^2}{2}\left( 1-z-2z^2-M +2Mz  \right) ;\nonumber \\
&&\dot{\bar{\pi}} +3 H \bar \pi +V_1=0 \,\, ; \quad V_1=\frac{dV(\phi)}{d \phi}\,; \quad \bar\pi\,\rm{has \,\, been\,\, defined\,\, as\,\,} \bar\pi= \dot \phi_0 ; \nonumber \\
&&\dot z= H \left(M-1-z\right ); \qquad \alpha_3=0=\alpha_4 ;.\nonumber \\
&& 1+z= \frac{b}{a} = \frac{H}{H_0}, \quad H_0 = \frac{\dot{b}}{M b}.
\eeq
 The algorithm one follows is similar to the one on \textit{page 5} with a few important additions:\\
1) Solve for $N_i$ and plug back in the Lagrangian;\\
\newline
2) Employ the field redefinitions:

\beq 
&&h_{ij}\rightarrow a^{1/2} h_{ij},\,\, \,\,p^{ij}\rightarrow 
a^{-1/2} p^{ij},\, \,\,\, n\rightarrow a^{-3/2} n ;\,\,\,\, \pi\rightarrow a^{3/2} \pi; \,\,\,\, \varphi\rightarrow a^{-3/2}\varphi;\,\,\,\, \nonumber \\ &&\bar{\Pi} \rightarrow a^3 \Pi;\,\,\,\, \p_i^2/a^2 \equiv \Delta.
\eeq
\newline
3) Solve for the $n$ constraint for the variable $p^t$ (no loss of generality);\\
\newline
4) Perform the field redefinitions:\\
 \beq
&&p^l \rightarrow p_0 +\frac{\nu^2}{4H} h^t,\, \,\,\, h^l\rightarrow q_0 +\frac{1}{2}h^t,\,\,\,\,             
\pi\rightarrow p_1-\frac{V^{'}}{4H}h_t, \nonumber \\&& \,\,\,\, \varphi \rightarrow q_1 + \frac{\bar{\Pi}}{4H}h_t,       \,\,\,\,\nu^2=m^2\,(1-z-2z^2)-2H^2+\frac{\bar{\pi}^2}{2}.
\eeq
\newline
5) Solve for $h^t$ by its algebraic equation;\\
\newline
6) Use the background e.o.m. and re-arrange the Lagrangian, through a series of canonical transformation, in such a way that it eventually takes the form:
\beq
L=P^{T}\cdot \dot{Q}-\Big[ \alpha P_0^2+ P_1^2 - \beta Q_0 \Delta Q_0 -\frac{Q_1 \Delta Q_1}{2}+ \delta\,\, Q_0 P_1 + \frac{Q_i \, M_{ij}\, Q_j}{2}\Big],  \label{finale1}
\eeq
 where one can prove \cite{Grisa} that the second to last term can be reabsorbed\footnote{We do not perform this step in the text because it is not necessary in order to determine the stability properties we are after. }. 
The canonical transformations necessary to this aim are three. It is quite easy to show that they are indeed canonical; it is also important to check one has enough freedom on the matrices that define these transformations as to be able to eventually put the Lagrangian in the form (\ref{finale1}).

The first transformation reads:
\beq
{\bf p}\to  {\bf P}+ h_{(,)} \cdot \left({\bf Q}+  \alpha_{(,)} \cdot{\bf P}\right),
\eeq
with $\alpha, h$ generic symmetric $2\times2$ matrices. These amounts to 6 degrees of freedom  in $\alpha,h$. 
The second transformation is simply: $Q_0\rightarrow -P_0,\,\,\, P_0 \rightarrow Q_0 $. After these transformations have been applied, one rewrites the Lagrangian in the form (\ref{forma}) and finds the entries of the new $K,M,V$ matrices. The requirements we impose on these new matrices stems from the form of Eq.~(\ref{finale1}):  one wants the (1,1) entry of the matrix $K$ to be one, the off diagonal elements to vanish and the (0,0) entry to be free of Laplacians\footnote{Note that we focus on K now because the second canonical transformation flipped momenta and coordinates.}. That leaves three free parameters. Those can be fixed by reproducing the wanted value of the coefficients multiplying $  Q_1 \Delta Q_1$, $Q_1 \Delta Q_0$ and $P_1 \Delta Q_0$, which saturates the number of free parameters and, at the same time, guarantees a Lagrangian in the form (\ref{finale1}). There is a third canonical transformation, which gives its desired result by construction (i.e. without fixing any of the free parameters). It reads:
\beq
{\bf P} \rightarrow {\bf P}+ A\cdot {\bf Q}, \quad {\bf Q}\rightarrow {\bf Q}, \qquad  {\rm with}\,\,  A= {\rm diag}\big[-V_{00}/K_{00}, -V_{11}/K_{11} \big]
\eeq 
This last step forces a   final $V$ matrix with  $V_{10}$ as the only non-zero entry.

 The Higuchi bound can now be read off directly from the sign of the $K_{00}$ entry. Quite remarkably, upon further simplifying the expression it turns out the $\dot{H}$-dependence drops out of the equations, to give a simple bound of the form: 
\
\beq
\frac{(3 H_0-2H) m^2 \Big[(3 H_0-2H ) m^2-2 H H_0^2\Big]}{12 H_0^4}\ge 0
\eeq
 where we have used: $\frac{\dot b}{M b}=H_0$ and $(1+z)=H/H_0$. As usual requiring the vector modes are stable restricts this further to
\be
(3 H_0-2H ) m^2-2 H H_0^2 \ge 0 .
\ee
This inequality immediately reduces to the usual $m^2\ge2H^2$ in the case without matter and with $f_{\mu\nu}=\bar{g}_{\mu\nu}$\footnote{It also gives back the result of \cite{Grisa} upon setting $\alpha_{3,4}$ to zero and using the appropriate background e.o.m.'s, i.e. $z=0$.}. 

There is no need to elaborate further on the $H>3/2H_0$ branch at this point, as the discussion would be identical to the one in \textit{Section 4}. As far as the $H<3/2H_0$ branch is concerned, one realizes that things are simply worsened by the realization that, the very same inequality $m^2\ge 2H H_0^2/(3H_0-2H)$ has now to hold over time. As we have seen before, making use of $\alpha_3, \alpha_4$ also does not change the qualitative picture. The general bound is (see \cite{paper2} for a full derivation)
\be
\tilde{m}^2(H) =m^2  \frac{H}{H_0} \left( (3+3\alpha_3+\alpha_4) -2(1+2 \alpha_3+\alpha_4)\frac{H}{H_0}+(\alpha_3+\alpha_4) \frac{H^2}{H_0^2}\right) \ge 2 H^2. \nonumber
\ee 
and is also independent of $\dot{H}$ and the precise form for matter. It might still be interesting to describe in some detail what happens if one picks some special values for $\alpha_3,\alpha_4$. A nice feature is that for $(\alpha_3+\alpha_4)=0$ the $z^3$ proportional contributions to the Friedmann equation and to the bound both vanish. At the same time, with $\alpha_3$ we have gained one parameter to tackle the Vainshstein screening mechanism: one can set the value of $\alpha_3=-1$ so that only $z$ appears in the Friedmann equation and that should make things easier for large $H$ (remember, $z=H/H_0 -1$). As it turns out after a quick calculation, the Higuchi bound, now easily satisfied upon choosing a small value for $H_0$,  forces $m^2/H_0 \gtrsim H$ which spoils the Vainshtein back again. This pattern will reappear again for other tempting values of $\alpha_3,\alpha_4$ and one essentially notices that together with the freedom that comes with the $\alpha_n$'s , there comes a polynomial in $z(t)$  which brings more conditions on the $\alpha_n$s in order for the Vainshstein mechanism to be at work.   

Although in some respects technically different from what the current literature offers\footnote{This is a different, better theory of MG, where the freedom on $f_{\mu\nu}$ is also fully employed. We have indeed seen how there is now an additional branch which does relax the Higuchi bound.},  once confronted with observations, the Higuchi bound we found still leaves little room if any for the value of $m$.


\section*{Conclusions}
\label{sec:Conclusions}
Recently, a new theory of massive gravity has been put forward \cite{deRham:2010kj}. Most interestingly, dRGT theory  is free from instabilities such as the Boulware-Deser ghost and tames issues which have plagued previous attempts at giving mass to the graviton. The Higuchi bound is a condition obtained directly from the study of the classical stability of a given theory. It is therefore crucial to investigate how dRGT is affected by this bound. We have performed such a study in this paper for the dRGT model with an FRW reference metric, which has revealed several, interesting things. First of all, once stability and observations are taken into account, the Higuchi bound is still quite stringent on the value of $m$, the parameter deforming GR. In summary the bound states that
\be
\tilde{m}^2(H) =m^2  \frac{H}{H_0} \left( (3+3\alpha_3+\alpha_4) -2(1+2 \alpha_3+\alpha_4)\frac{H}{H_0}+(\alpha_3+\alpha_4) \frac{H^2}{H_0^2}\right) \ge 2 H^2. \nonumber
\ee 
where $H$ is the Hubble rate of the dynamical metric and $H_0$ that of the reference metric.

This bound differs significantly from previous conditions, in particular \cite{Grisa} (see also \cite{Blas,Berkhahn:2010hc,Berkhahn:2011hb}) in that it is independent of $\dot{H}$ and the precise form of matter. Indeed the bound found in \cite{Grisa} $m^2 > 2(H^2 +\dot{H}/N)$ appears to allow without constraint accelerating solutions. By contrast the bound found for the dRGT models is equally strong regardless of whether the universe is in an accelerating or decelerating regime. Furthermore, unlike these previous works, here we have studied a theory where the reference metric is distinct from the background dynamical metric. 

Generically when analyzing the Higuchi inequality, one identifies two or more branches of values for the Hubble parameter $H$ that satisfy the condition. For the case $\alpha_3=\alpha_4=0$, one branch contains the $f_{\mu\nu}=\bar{g}_{\mu\nu}$ solution in the absence of matter while the other one allows for a negative $m^2$ value whilst the effective mass $\tilde{m}^2$, the one that counts, is kept positive. If one were to focus just on the instability at this stage, it would be correct to conclude that the branch which does not include $f_{\mu\nu}=\bar{g}_{\mu\nu}$ does indeed allow for a very weak, easily satisfied, Higuchi bound. However, one soon realizes this set of values for the parameters of the theory does not stand the test of observations, i.e. recovering GR wherever the Vainshtein screening mechanism is at work.

Another feature that emerged in this work is that the free parameters $\alpha_3, \alpha_4$ do not change the qualitative picture of the Higuchi-Vainshtein analysis; they would if one were frozen into a specific cosmological epoch, but fully considering that $H=H(t)$ and that one must account for all type of matter, is enough to render the coefficients ineffective to this aim. 

It is interesting to note at this point how the analysis of homogeneous and isotropic cosmological solutions in dRGT \cite{massivec} might carry the same message we learned here. In that work, a no-go theorem for $k=0$ FRW solutions to dRGT with Minkowski reference metric was presented. The way out there relied on recalling that observations only require approximate homogeneous and isotropic  solutions, and on looking for them in dRGT \cite{massivec, Gratia:2012wt,Kobayashi:2012fz,Volkov:2012cf,DeFelice:2012mx,Gumrukcuoglu:2012aa}. We plan to lift the purely homogeneous and isotropic assumptions on the metrics in a follow up paper \cite{paper3}.  As an additional note it should be said that the no-go theorem can be evaded by giving full dynamics to both metrics, that is considering multi-metric theories \cite{Pilo,vonStrauss,Hassan:2011zd,Hinterbichler:2012cn,Hassan:2012wt}. We did not completely rule out here the fact that multimetric theories could also help in relaxing the Higuchi-Vainshtein tension and shall return to this in subsequent work \cite{paper2}. 

Verifying that dRGT theory possesses stable regions of the parameters space where the Higuchi-Vainshtein tension can be relaxed is quite an important task. We have shown that this is not the case when both the reference and dynamical metrics are FRW. We will report on our the results on the possible resolutions in a forthcoming work \cite{paper2,paper3}. 

Whilst this paper was being submitted, the following paper \cite{Berg} appeared in the literature. The study in \cite{Berg} very partially overlaps with the work presented here.  The authors there focus on the de Sitter and quasi-de Sitter case (see our \textit{Section 4}) for bigravity. They do not study general FRW backgrounds as we do here. Also \cite{DeRham} appeared in the literature at the same time. There the authors study the so called decoupling limit of solutions such as the ones described in this paper, with special attention to the Higuchi bound and Vainshtein mechanism in that limit.

\acknowledgments

We would like to thank Claudia de Rham, David Langlois and Atsushi Naruko for valuable remarks and conversations. AJT and MF would like to thank the  Universit\'e de  Gen\`eve for hospitality while part of this work was being completed. AJT is supported by DOE grant DE-FG02-12ER41810.

\section{Appendix \textit{A}}
The explicit expression for the functions $c_{n},d_n,e_n,f_n$ in the MG action of Eq.~(\ref{semiesplicita}) reads:
\beq
&&c_2= \left(\frac{3}{4 a}-\frac{3 b}{8 a^2}-\frac{3 M}{8 a}+\frac{b M}{8 a^2}\right);\quad d_2= \left(-\frac{3}{2 a}+\frac{9 b}{8 a^2}-\frac{b^2}{8 a^3}+\frac{3 M}{4 a}-\frac{3 b M}{8 a^2}\right);\nonumber \\ && e_2=\left(3 a-3 b+\frac{b^2}{2 a}\right); \quad f_2= \left(\frac{3 b^2}{2 (b+a M)}-\frac{b^3}{a (b+a M)}\right);\nonumber \\&&
c_3=  \left(-\frac{1}{2 a}+\frac{3 b}{8 a^2}+\frac{3 M}{8 a}-\frac{b M}{4 a^2}\right); \quad
d_3= \left(\frac{1}{a}-\frac{9 b}{8 a^2}+\frac{b^2}{4 a^3}-\frac{3 M}{4 a}+\frac{3 b M}{4 a^2}-\frac{b^2 M}{8 a^3}\right);\nonumber \\ &&
e_3= \left(-2 a +3 b -\frac{b^2 }{a}\right);\quad  f_3=\left(-\frac{3 b^2}{2 (b+a M)}+\frac{2 b^3}{a (b+a M)}-\frac{b^4}{2 a^2 (b+a M)}\right); \nonumber \\&&
c_4=  \left(\frac{1}{8 a}-\frac{b}{8 a^2}-\frac{M}{8 a}+\frac{b M}{8 a^2}\right);\quad d_4= \left(-\frac{1}{4 a}+\frac{3 b}{8 a^2}-\frac{b^2}{8 a^3}+\frac{M}{4 a}-\frac{3 b M}{8 a^2}+\frac{b^2 M}{8 a^3}\right);\nonumber \\ &&
e_4= \left(\frac{a }{2}-b +\frac{b^2 }{2 a}\right); \quad f_4=\left(\frac{b^2}{2 (b+a M)}-\frac{b^3}{a (b+a M)}+\frac{b^4}{2 a^2 (b+a M)}\right) .
\eeq
 \newline\\
The explicit expression for the quantity $A$ in Eq.~(\ref{canonical1}) reads:
\beq
\frac{2 H \left(-9 H^2-3 m^2 (1+z) (-1+z (2-\alpha_3 (-2+z)-\alpha_4 z))-2 \Delta \right)}{m^2 (1+z) (-1+z (2-\alpha_3 (-2+z)-\alpha_4 z)) } \cdot \nonumber  \\ \times \frac{1}{\left(2 H^2+m^2 (1+z) (-1+z (2-\alpha_3 (-2+z)-\alpha_4 z))\right)}
\eeq

\section{Appendix \textit{B}}
We give below the entries of the matrices $K, M, V$ as they appear before point (6) in \textit{Section 5}, that is, before the three final  canonical transformations are performed:

\beq
 \mathbf{K}_{\mu\nu} = \left(
      \begin{array}{ccc}
        3+ \frac{4H (1+z)+2 \dot z}{H m^2 (1+z)^2 (-1+2z)} \Delta - \frac{4(H+H z + \dot z)}{3 H m^2(1+z)\nu^2} \Delta^2 & \quad  \frac{\bar \pi}{2H }- \frac{  \bar{\pi}(H+ H z + \dot z)  }{3 H^2 (1+z)\nu^2} \Delta \\
        \frac{\bar \pi}{2H }- \frac{  \bar{\pi}(H+ H z + \dot z)  }{3 H^2 (1+z)\nu^2} \Delta   & 1+ \frac{m^2 \bar{\pi}(1-2z)(H+H z +\dot z)}{12 H^3 \nu^2}
      \end{array} \right)\,\,\, ,  \qquad \quad \,\,\, \label{kappa11}
\eeq

\beq
 \mathbf{V}_{00} =
    &&    -\frac{5}{2}H -\frac{\nu^2}{2H}+\frac{\Delta}{3 H}+\frac{4 H \Delta}{m^2(1-z-2z^2)}+\frac{\dot z \Delta}{3 H^2 (1+z)}-\frac{2 \dot z \Delta}{m^2(1+z)^2 (-1+2z)}\nonumber \\
&&-\frac{4 H \Delta^2}{3m^2 \nu^2 (1-z-2z^2)}+\frac{4 \dot z \Delta^2 }{3m^2 \nu^2 (1+z)^2(-1+2z)} 
\eeq
\beq 
 \mathbf{V}_{01} =&&\frac{V_1}{2H}+\frac{2 \bar{\pi} \Delta}{m^2(1-z-2z^2)}-\frac{V_1 \Delta}{3H \nu^2}-\frac{\bar{\pi} \dot{z} \Delta}{H m^2 (1+z)^2(-1+2z)}-\frac{v_1 \dot{z} \Delta }{3H^2 \nu^2 (1+z)}\,\, \nonumber \\
&&-\frac{2 \bar{\pi} \Delta^2 }{3 m^2 \nu^2 (1-z-2z^2)}+\frac{2 \bar{\pi} \dot{z} \Delta^2}{3H m^2 \nu^2 (1+z)^2(-1+2z)}
\eeq
 
\beq
\mathbf{V}_{10}= -\bar{\pi}-\frac{m^2(1-z-2z^2)\bar{\pi}}{12 H^2}+\frac{m^2 \bar{\pi} (-1+2z) \dot{z}}{12 H^3}+\frac{\bar{\pi} \Delta}{3\nu^2}+\frac{\bar{\pi} \dot{z} \Delta}{3H \nu^2 (1+z)}\qquad \quad 
\eeq
\beq
\mathbf{V}_{11}=\frac{m^2 \bar{\pi} V_1 (1-2z) \dot{z} }{12 H^3 \nu^2 }+\frac{3}{2 H}+\frac{m^2 \bar{\pi} V_1 (1-z-2z^2)}{12 H^2 \nu^2 }+\frac{\bar{\pi}^2 \Delta }{6 H \nu^2 }+\frac{\bar{\pi}^2 \dot {z} \Delta}{6 H^2 \nu^2 (1+z)} \nonumber \\
\eeq

\beq
&&\mathbf{M}_{00}= \frac{1}{24 H^3}\left(H \left(4 H^2 \left(6 \bar{\pi}^2-5 m^2 \left(-1+z+2 z^2\right)\right)+m^2 (1+z) (-1+2 z)  \right. \right. \times\nonumber \\
&& \left. \left. \left(-\bar{\pi}^2+2 m^2 \left(-1+z+2 z^2\right)\right)\right)+m^2 (-1+2 z) \left(16 H^2-\bar{\pi}^2+2 m^2 \left(-1+z+2 z^2\right)\right) \dot{z}\right)\nonumber \\
&&
-\frac{2 \left(-6 H^3 (1+z)+H m^2 (1+z)^2 (-1+2 z)-3 H^2 \dot{z}+m^2 \left(-1+z+2 z^2\right) \dot{z}\right) \Delta }{3 H m^2 (1+z)^2 (-1+2 z)}\nonumber \\ &&-\frac{4 H (H+H z+\dot{z})  \Delta^2   }{3 m^2 (1+z)^2 (-1+2 z) \nu^2 }
\eeq

\beq
&&\mathbf{M}_{01}=\mathbf{M}_{10}= \frac{m^2 V_{1} (-1+2 z) (H+H z+\dot{z})}{12 H^3}-\frac{ \bar{\pi} (H+H z+\dot{z}) \Delta }{6 H^2 (1+z)}+\frac{ V_{1} (H+H z+\dot{z}) \Delta }{3 H (1+z) \nu ^2}\nonumber \\ &&+\frac{ \bar{\pi} (2 H (1+z)+\dot{z}) \Delta }{ m^2 (1+z)^2 (-1+2 z)}-\frac{2 \bar{\pi} (H+H z+\dot{z}) \Delta ^2}{3 m^2 (1+z)^2 (-1+2 z) \nu ^2}
\eeq

\beq
&&\mathbf{M}_{11}= \frac{H \left(6 H^2 \left(-4 H^2+\bar{\pi}^2\right) V_{2}-m^2 \left(V_{1}^2+12 H^2 V_{2}\right) \left(-1+z+2 z^2\right)\right)+m^2 V_{1}^2 (1-2 z) \dot{z}}{12H^3 \nu^2}\nonumber \\ &&
-\Delta -\frac{2 \bar{\pi} V_{1} (H+H z+\dot{z})}{H^2 (1+z) 3\left(4 H^2-\bar{\pi}^2+2 m^2 \left(-1+z+2 z^2\right)\right)}\Delta +\frac{ \bar{\pi}^2 (2 H (1+z)+\dot{z})}{2 H m^2 (1+z)^2 (-1+2 z)}\Delta  \nonumber \\ &&-\frac{ \bar{\pi}^2 (H+H z+\dot{z}) \Delta ^2}{3H m^2 (1+z)^2 (-1+2 z)\nu^2} \label{emme1}.
\eeq
 Starting with generic (up to this point) entries for the  symmetric matrices $\alpha$ and $h$, we show their final formal expression after the three canonical transformations are performed. We further show how to determine the various entries by the requirement that the final Lagrangian is of the form (\ref{finale1}). First, the generic symmetric matrices:
\beq
 \mathbf{h}_{a\,b} = \left(
      \begin{array}{ccc}
        h_{00} &h_{01} \\
       h_{01}   & h_{11}
      \end{array} \right)\,\,\,  \qquad \quad \,\,\,
 \mathbf{ \alpha }_{c\,d} = \left(
      \begin{array}{ccc}
       \alpha_{00} &  \alpha_{01} \\
       \alpha_{01}   &  \alpha_{11}
      \end{array} \right)\,\,\,  \qquad \quad \,\,\,
\eeq
After the following canonical transformations
\beq
&&\left(
\begin{array}{c}
 p _0\\
 p_1 \\
\end{array}
\right)\rightarrow
\left(
\begin{array}{c}
 P_0 \\
 P_1 \\
\end{array}
\right)+\left(
\begin{array}{cc}
 h_{00} & h_{01} \\
 h_{01} & h_{11} \\
\end{array}
\right).\left(
\begin{array}{c}
 Q_0 \\
 Q_1 \\
\end{array}
\right)+\left(
\begin{array}{cc}
 h_{00} & h_{01} \\
 h_{01} & h_{11} \\
\end{array}
\right).\left(
\begin{array}{cc}
 \alpha_{00} & \alpha_{01} \\
 \alpha_{01} & 0 \\
\end{array}
\right).\left(
\begin{array}{c}
 P_{0} \\
 P_{1} \\
\end{array}
\right)
\nonumber \\
&&\left(
\begin{array}{c}
 q_{0} \\
 q_{1} \\
\end{array}
\right)\rightarrow
\left(
\begin{array}{c}
 Q_{0} \\
 Q_{1} \\
\end{array}
\right)+\left(
\begin{array}{cc}
 a_{00} & a_{01} \\
 a_{01} & 0 \\
\end{array}
\right).\left(
\begin{array}{c}
 P_{0} \\
 P_{1} \\
\end{array}
\right),
\eeq
\beq
P_0\rightarrow Q_0,\qquad  \qquad Q_0\rightarrow -P_0,
\eeq
 one has to write the Lagrangian in the form (\ref{forma}) according to the new variables $({\bf Q} ,{\bf P})$. This, after many integrations by part, will give the new $K,V,M$ matrices. As an example we report below the new $K$ matrix.
\beq
\tilde {\bf K}_{00}=&&\dot h_{00} + h_{00}^2 k_{00} + h_{01}^2 k_{11} + M_{00} + 2 h_{00} (h_{01} K_{01} + V_{00}) + 
 2 h_{01} V_{10}, -h_{00} K_{01} - h_{01} K_{11} - V_{10}\nonumber \\ &&- 
 \alpha_{01} (\dot h_{00} + h_{00}^2 K_{00} + h_{01}^2 K_{11} + M_{00} + 2 h_{00} (h_{01}  K_{01} + V_{00}) + 
    2 h_{01} V_{10})\nonumber \\
\tilde {\bf K}_{01}=&&-h_{00} K_{01} - h_{01} K_{11} - V_{10} - 
 \alpha_{01} (\dot h_{00} + h_{00}^2 K_{00} + h_{01}^2 K_{11} + M_{00} \nonumber \\ &&+ 2 h_{00} (h_{01} K_{01} + V_{00}) + 
    2 h_{01} V_{10})\nonumber \\
\tilde {\bf K}_{11}=&&K_{11} + 2 \alpha_{01} (h_{00} K_{01} + h_{01} K_{11} + V_{10}) + 
 \alpha_{01}^2 (\dot h_{00} + h_{00}^2 K_{00} + h_{01}^2 K_{11} + M_{00} \nonumber \\ &&+ 2 h_{00} (h_{01} K_{01} + V_{00}) + 
    2 h_{01} V_{10})
\eeq
 where the matrix entries $K,M,V$ on the RHS of this last equation are to be intended as the ``original'' $K,M,V$ entries, as given in Eq.~(\ref{kappa11}-\ref{emme1}) above.
Equipped with the new matrices one then requires:\\
- $\tilde {\bf K}_{01}=0$, and solves for $\alpha_{01}$;\\
- $\tilde {\bf V}_{01}=0$ and solves for $h_{11}$;\\
- $\tilde {\bf K}_{11}=1$ and solves for $h_{01}$;\\
This is clearly taking the Lagrangian towards the form in (\ref{finale1}). The other conditions that  will further fix the remaining entries (e.g. $h_{00}=H_{Hubble}$, etc.) are:\\
- No $\Delta$'s in $\tilde {\bf K}_{00}$ ;\\
- No $\Delta$'s in $\tilde {\bf M}_{01}$ ;

One can further massage the final expression for the Lagrangian, but, after this point, further requirements on $\alpha, h $ entries must be true by construction (i.e. the requirement will be a consequence of the ones above)  because the freedom on the matrices has been saturated. We give the explicit expression for the final form of the $\tilde {\bf K}_{00}$ entry in the text (\textit {Section 5}) as it is by requiring it to be positive that one obtains the Higuchi bound.



\end{document}